\documentclass[11pt]{article}

\usepackage[nottoc,notlot,notlof]{tocbibind}

\usepackage[T1]{fontenc}
 \usepackage[english]{babel}
\usepackage{layout}
\usepackage{natbib}
\usepackage{amsmath}
\usepackage{mathrsfs}
\usepackage{amssymb}
\usepackage{changes}
\usepackage{graphicx}
\usepackage[euler]{textgreek}
\usepackage{ucs}
\usepackage{float}
\usepackage{lipsum}
\usepackage{listing}
\usepackage{bibentry}
\usepackage{appendix}
\usepackage{upgreek}
\usepackage{chngcntr}
\usepackage{listings}
\usepackage[makeroom]{cancel}
\usepackage{newfloat}
\usepackage{setspace}
\usepackage{enumitem}
\usepackage{cite}

\usepackage{bbm, dsfont}
\usepackage{tcolorbox}
\usepackage{textcomp}
\usepackage{lineno}
\usepackage[colorlinks=true,citecolor=blue,linkcolor=blue]{hyperref}%
\usepackage[a4paper,top=2cm, bottom=2cm, left=2cm, right=2cm]{geometry}
\usepackage{todonotes}
\usepackage{siunitx}
\usepackage{caption}
\usepackage[ruled,vlined]{algorithm2e}
\usepackage{relsize} 

\usepackage{ragged2e}
\usepackage[scaled=.7]{beramono}
\usepackage{pdfpages}
\usepackage{authblk}
\usepackage{cancel}
\usepackage{newfloat}
\DeclareFloatingEnvironment[name={Appendix 1-Figure}]{suppfigure}
\DeclareFloatingEnvironment[name={Appendix 1-Table}]{supptable}

\usepackage[capitalize,noabbrev]{cleveref}
\crefformat{figure}{#2\figurename~#1#3}
\crefformat{table}{#2\tablename~#1#3}
\crefname{suppfigure}{Appendix 1-Figure}{Appendix 1-Figures}
\crefname{supptable}{Appendix 1-Table}{Appendix 1-Tables}
\crefformat{suppfigure}{#2\suppfigurename~#1#3}
\crefformat{supptable}{#2\supptablename~#1#3}

\definechangesauthor[color=cyan]{JJ}
\definechangesauthor[color=brown]{ND}
\definechangesauthor[color=red]{WS}
\definechangesauthor[color=pink]{MAP}
\definechangesauthor[color=violet]{R}

\newcommand{\x}{\boldsymbol{x}}

\newcommand{\z}{\boldsymbol{z}}

\makeatletter

\title{Learning beyond sensations: how dreams organize neuronal representations}
\author{Nicolas Deperrois$^{1}$\footnote{Correspondence: nicolas.deperrois@unibe.ch}~, Mihai A. Petrovici$^{1}$, Walter Senn$^{1}$, and Jakob Jordan$^{1,2}$}
\affil{$^{1}$Department of Physiology, University of Bern, Bern, Switzerland}
\affil{$^{2}$Electrical Engineering, Yale University, New Haven, CT, United States}

\date{}

\begin{document}

\maketitle

% external
%  - sensations, sensory input, sensory stimuli, externally driven activities
% internal
%  - virtual experiences, imagined experiences

%%%%%%%%%%%%%%%%%%%%%%%%%%%%%%%%%%%%

% ABSTRACT

%%%%%%%%%%%%%%%%%%%%%%%%%%%%%%%%%%%%

\begin{abstract}
    Semantic representations in higher sensory cortices form the basis for robust, yet flexible behavior.
    These representations are acquired over the course of development in an unsupervised fashion and continuously maintained over an organism's lifespan.
	Predictive processing theories propose that these representations emerge from predicting or reconstructing sensory inputs. 
	However, brains are known to generate virtual experiences, such as during imagination and dreaming, that go beyond previously experienced inputs.
	Here, we suggest that virtual experiences may be just as relevant as actual sensory inputs in shaping cortical representations.
	In particular, we discuss two complementary learning principles that organize representations through the generation of virtual experiences.
	First, ``adversarial dreaming'' proposes that creative dreams support a cortical implementation of adversarial learning in which feedback and feedforward pathways engage in a productive game of trying to fool each other.
	Second, ``contrastive dreaming'' proposes that the invariance of neuronal representations to irrelevant factors of variation is acquired by trying to map similar virtual experiences together via a contrastive learning process.
	These principles are compatible with known cortical structure and dynamics and the phenomenology of sleep thus providing promising directions to explain cortical learning beyond the classical predictive processing paradigm. 
	% We additionally discuss the computational benefits of such principles beyond learning semantic representations, such as enhancing creative insight, and their neuronal and behavioral correlates. 
\end{abstract}

%%%%%%%%%%%%%%%%%%%%%%%%%%%%%%%%%%%%

% TABLE OF CONTENTS

%%%%%%%%%%%%%%%%%%%%%%%%%%%%%%%%%%%%

%\tableofcontents

%%%%%%%%%%%%%%%%%%%%%%%%%%%%%%%%%%%%

% INTRODUCTION

%%%%%%%%%%%%%%%%%%%%%%%%%%%%%%%%%%%%

\section{Introduction}

% make the reader aware of contrast: unique sensory experiences, but reliable behaviour
Throughout their life, animals enjoy a wide variety of unique sensory experiences.
However, seemingly unaffected by this diversity, animals exhibit a remarkable degree of consistency in their behaviour and can, often effortlessly, leverage prior knowledge to generalize to novel circumstances.
For example, they easily recognize which category an object belongs to \citep{biederman1987recognition}, within a fraction of a second \citep{thorpe1996speed}, and despite the various conditions in which this object can be observed \citep{dicarlo_how_2012}.
How is this possible?

\begin{figure}[t!]
   \centering
   \includegraphics[width=1.0\textwidth]{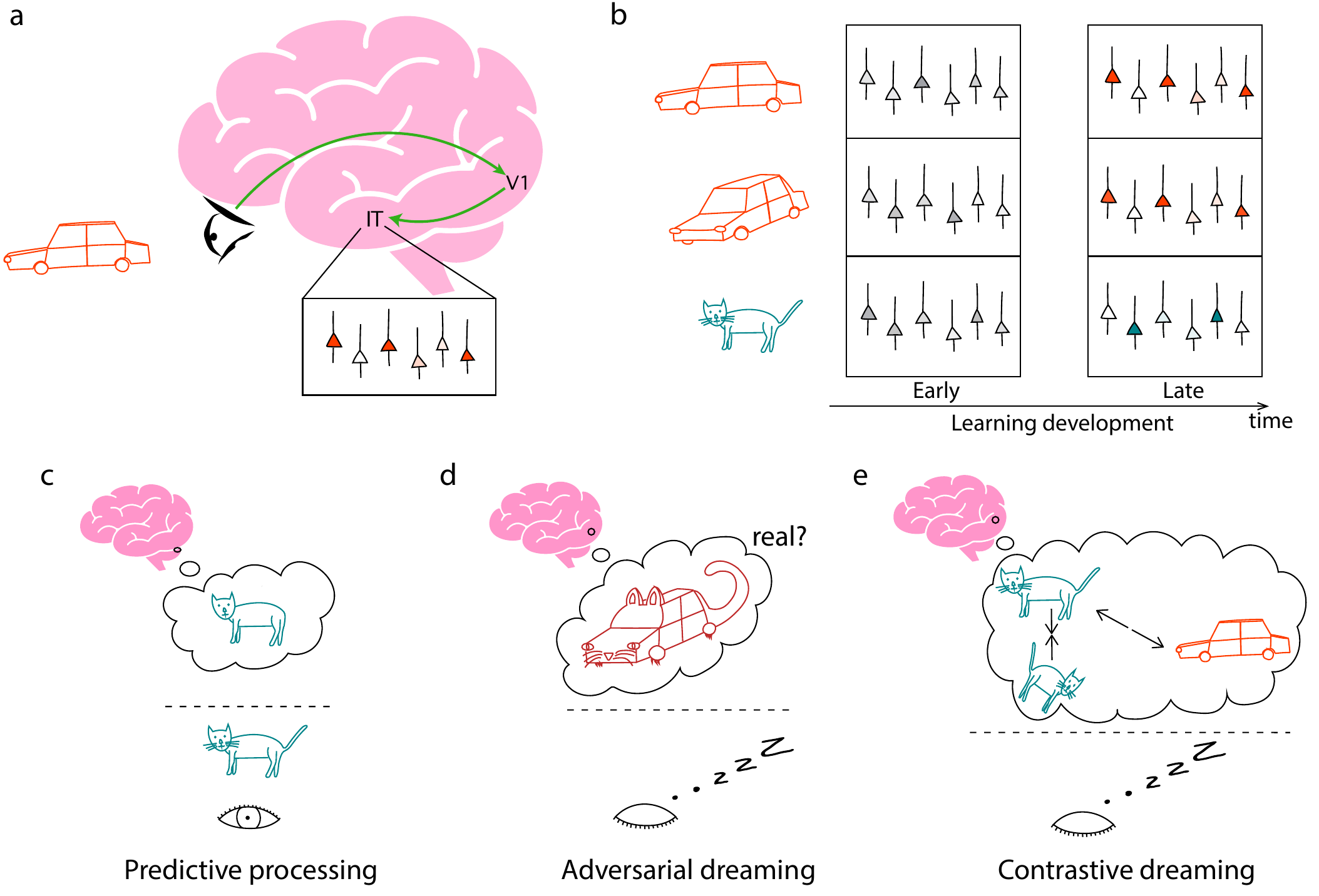}
   \caption{{\bf Semantic representations in higher cortical areas emerge over the course of development.}
   \small 
   \textbf{(a)} Sketch of time-averaged activity of neurons in IT in response to a visual stimulus. Visual stimuli activate cells in the retina and these signals are processed along the hierarchy of the visual cortex, here the ventral visual stream.
   \textbf{(b)} Typical neuronal responses to the presentation of different objects at early and late stages of development. Over the course of development, activity patterns align with the semantic category of the input (here, different patterns encode ``cat'' and ``car'' stimuli) but are invariant to semantically preserving transformations (e.g., cars from different viewing angles).
   \textbf{(c)} Common neuroscientific theories hypothesize that brains learn representations by trying to predict their sensorium \citep[predictive processing,][]{rao_predictive_1999}.
   \textbf{(d)} During offline states, e.g., sleep, brains continue to generate virtual experiences that may further contribute to learning semantic representations, for example by combining several memories into new, realistic experiences. 
   \textbf{(e)} In addition, regenerating previous experiences with natural semantically preserving variations can provide additional signals for learning.
   }\label{fig:representation}
\end{figure}

% solution: semantic representations
Insights from neuroscience and machine learning suggest that this cognitive feat may be grounded in neuronal activity patterns in higher cortical areas that reflect the {\it semantic} content of sensory inputs.
Thus, these ``semantic neuronal representations'' extract relevant factors of variation such as object categories from stimuli while remaining invariant to irrelevant factors such as pose, lighting, or partial occlusions \citep{barlow2001redundancy,dicarlo_how_2012}.
Strikingly, such an organized and invariant code is observed in recordings from the inferior temporal (IT) cortex, the highest area of the ventral visual stream \citep[\cref{fig:representation}a;][]{grill-spector_2001_the, hung_fast_2005}.

% next contrast: semantic representations emerge, but not through labels
Such structure in neuronal activities arises over the course of development \citep[\cref{fig:representation}b;][]{rodman1994development}.
Yet, the mechanisms underlying this emergence remain unclear.
Computational models of the sensory cortex attempt to explain how, from sensory evoked activities (activities from lower cortical areas, e.g., in V1 cortex), neurons extract features of increasing complexity along the cortical hierarchy \citep{hubel1965receptive}, leading to high-level semantic representations \citep{richards_2019}.
Supervised models of sensory processing suggest that cortical feedforward pathways learn to map sensory inputs to specific object categories that are externally provided, for example by a teacher.
However, animals seem to learn with little to no supervision and do not require millions of category labels during development \citep{bergelson_at_2012, bergelson_nature_2017, slone_infants_2015, huber2021a}.

% solution: unsupervised learning
To acquire semantic latent representations, cortical networks may thus leverage learning principles that do not rely on labelled data, similar to unsupervised machine learning models \citep{liu_self-supervised_2021, zhuang_unsupervised_2021}.
For example, cortical feedback pathways could implicitly learn the structure of the sensorium by generating activities in lower sensory cortex that are similar to sensory-evoked responses \citep[generative modeling, ][]{clark_whatever_2013}.
Therefore, we first discuss the predictive processing framework  \citep[\cref{fig:representation}c,][]{rao_predictive_1999}, that posits that the brain shapes its representations by trying to predict evoked activities in early sensory cortex. We then present ``adversarial dreaming'' \citep[\cref{fig:representation}d;][]{deperrois_learning_2022} where representations are improved through the generation of creative dreams during sleep and their discrimination from actual experiences.

Another possibility is to directly shape feedforward pathways to construct relevant high-level representations, by using simple, alternative labels that can directly be inferred from data \citep[self-supervised learning, ][]{ericsson_selfsupervised_2022}. Accordingly, we introduce ``contrastive dreaming'' (\cref{fig:representation}e), during which neuronal activities in higher cortical areas are pulled together for semantically similar inputs, and pushed apart for dissimilar stimuli.

For each framework, we start by presenting the underlying computational principles and then discuss suggested bio-plausible implementations.
Finally, we discuss experimental approaches to (in)validate the presented hypotheses.

%%%%%%%%%%%%%%%%%%%%%%%%%%%%%%%%%%%%

% PREDICTIVE LEARNING 

%%%%%%%%%%%%%%%%%%%%%%%%%%%%%%%%%%%%

\section{Learning by predicting evoked low-level cortical activities}

%%% box 

\subsection{Principles of predictive processing}

% why prediction?
In perception, an efficient way to represent relevant aspects of the sensorium is to try to ``explain away'' evoked activities in lower sensory cortex through a cascade of predictions performed by cortical feedback pathways \citep{helmholtz1878facts, clark_whatever_2013}.
These predictions reflect what the brain already knows about the sensorium. Informally, one may consider them emerging from a set of priors acquired through experience.
By trying to match the sensory evoked signal, the brain thus seeks for latent causes that would best characterize the stimulus, such as its semantic category.
Through these processes, brains can learn organized semantic representations \citep{friston_free-energy_2010, clark_whatever_2013}.

These ideas have been formalized by computational frameworks such as predictive coding \citep[\cref{fig:representation}c;][]{rao_predictive_1999}, which describes how neuronal dynamics and synaptic plasticity are both involved in learning generative models of the environment.
%
% % predictive coding
First, on short time scales, neuronal activities change to better predict the evoked activity.
Second, on longer time scales, synaptic plasticity aims to further improve these predictions (Box 1).
Over time, minimization of prediction errors through these dynamics implicitly organizes latent representations \citep{lotter2017deep}.
In computational models, the prediction error is usually a point-wise measure at an appropriately coarse graining of the stimulus.
For example, for images, the prediction error is typically measured at a pixel level.
In the brain, it is thought to be computed by subclasses of layer 2/3 pyramidal neurons
\citep{mumford1992computational, bastos2012canonical, shipp2016neural} that compare activities between predictive neurons and neurons activated by sensory inputs \citep{Keller2018}. 

\medskip
\begin{tcolorbox}
\textbf{Box 1 - Predictive processing theories} 

Predictive coding \citep{rao_predictive_1999} hypothesizes that brains minimize errors between predictions generated by feedback pathways and low-level evoked activities. This serves two complementary functions: finding latent activities that are compatible with a presented stimulus (inference) and adjusting synaptic weights to improve predictions (learning).
	
Mathematically, predictive coding can be described as a special case of variational inference \citep{jordan1999introduction,friston_theory_2005,marino_predictive_2022}. Assuming a Gaussian distribution for the generative model and approximate posterior, predictive coding infers a latent activity $\z^*$ that (locally) minimizes the following loss via gradient descent
\begin{align}
	\mathcal{L_{\text{pred}}} =  \| \x - G(\z) \|^2 + \| \z - \boldsymbol{\mu_z}\|^2 \, ,
	\label{eqn:predictive_coding_inference}
\end{align}
where $\x$ represents sensory input, $\z$ the latent activity of the network, $G$ a (deep) generative network and $\boldsymbol{\mu_z}$ the mean of the latent prior.
Intuitively, predictive coding thus infers latent activities $\z^*$ by minimizing the reconstruction error between the generated prediction $G(\z)$ and the actual sensory input $\x$ (first term of \cref{eqn:predictive_coding_inference}).
The second term reflects the prior and can be interpreted as a regularization term that can implement activity constraints, such as sparsity \citep{rao_predictive_1999}.

Once latent activities are inferred, a gradient step with respect to the parameters of the generator $G$ is taken on \cref{eqn:predictive_coding_inference}, further reducing the reconstruction error between actual inputs $\z$ and predicted inputs $G(\z)$. 
These separate optimization steps assume that synaptic weight changes (and thus, learning) occur on a slower timescale than inference.
\end{tcolorbox}

% great, stuff, but...
Due to the hierarchical nature of the predictive coding framework, neuronal populations develop selectivity for low-level details in early areas and for high-level properties (objects, shapes, scenes) in later areas, compatible with experimentally measured neuronal responses and receptive fields \citep{rao_predictive_1999}. 
Moreover, building models based on these principles leads to the emergence of latent representations suitable for efficiently learning downstream tasks \citep{rumelhart1986learning, kingma2013autoencoding,lotter2017deep,millidge_predictive_2022}. 
Predictive processing principles thus suggest a computational model compatible with cortical structure and activity for learning in brains \citep[but see][]{koch1999predicting,murray2004perceptual} from a simple goal:~predicting sensory-evoked low-level cortical activities.

%%% Shortcomings of PC 

\subsection{Beyond the prediction of sensations}

As soon as we reduce external sensory inputs, for instance through unfocusing our eyes, meditation, or deep rest, we can become aware of virtual experiences continuously produced by our brain \citep{mildner2019spontaneous}.
These manifest in their strongest form as dreams, mostly occurring during the rapid-eye-movement (REM) phase of sleep \citep{nir_dreaming_2010}.
While dreams may feel familiar, they often represent objects, scenes, situations that go beyond what we previously experienced \citep{fosse_dreaming_2003, wamsley_dreaming_2014}.
Indeed, during REM dreams previous waking experiences are often not identically recalled but rather incorporated with other past memories into a new narrative \citep{fogel2018novel, northoff2023topographic}. 

The predictive processing framework has been previously suggested to also account for the phenomenology of virtual experiences during sleep \citep{hobson_waking_2012, hobson_virtual_2014}.
Accordingly, the same feedback pathways employed to generate predictions of sensory inputs during wakefulness, are ``freed'' from sensory inputs during sleep, allowing the generation of virtual experiences.
This has been proposed to contribute to minimizing the generative model's complexity, i.e., the degrees of freedom required to describe the sensorium, for example by pruning redundant synapses.
Consequently, dreaming would facilitate the ability to generalize and understand the semantics of the external world.

Here, we suggest that learning from virtual experiences has additional roles besides minimizing complexity.
As we will describe in the following, learning during offlines states can further improve the predictive model by increasing its realism, can sharpen our ability to distinguish between internally generated and externally driven activities, and robustify semantic neuronal representations against perturbations, such as occlusions of parts of the visual field.
To this end we discuss two complementary approaches, ``adversarial dreaming" and ``contrastive dreaming", which, as we will explain, are crucial for organizing neuronal representations.

%%%%%%%%%%%%%%%%%%%%%%%%%%%%%%%%%%%%

% ADVERSARIAL LEARNING 

%%%%%%%%%%%%%%%%%%%%%%%%%%%%%%%%%%%%

\section{Learning representations by creating virtual experiences}
	
\subsection{Principles of adversarial learning}

Cortical models that aim to also learn from virtual experiences, such as dreams, cannot rely exclusively on sensory prediction errors due to the absence of ground-truth evoked activities, but have to find alternative sources of learning signals.
For example, such models could learn to produce data that appears "similar" to previous stimuli without trying to exactly reproduce them.
But how can we quantify this similarity to derive useful learning signals?
One possibility is to expand the model with an additional module that learns to measure the similarly between generated data and actual stimuli.
In this spirit, Generative Adversarial Networks \citep[GANs,][]{goodfellow2014generative} introduce an architecture that consists of two networks:~a generator producing virtual samples and a discriminator judging whether a sample is real or generated.
These two networks are trained adversarially, with the discriminator learning to distinguish generated from real samples while the generator learns to fool the discriminator by improving the realism of the generated samples (Box 2). 
Through this adversarial game, the generator gradually learns to synthesize samples that are similar to the training data. 
This process can be illustrated by a student (generator) that tries to fake their parents' handwriting, and the teacher (discriminator) that detects whether the writing is real or fake. After many attempts, the student learns how to fool their teacher by writing in a way that is hardly distinguishable from their parents' handwriting.

The goal of this process is to find a balanced optimum in which the generator produces a large variety of samples with sufficient similarity to real samples.
For example, in the context of natural image generation, generated samples contain colors, shapes, and objects that are typically present in real images. 
However they can also be distorted or combined versions of these objects, as a consequence of adversarial learning, as even after convergence the generator keeps the freedom to generate creative samples \citep{brock_large_2019}.

Furthermore, GANs are known to extract semantic latent representations from data \citep{radford_unsupervised_2015, donahue_adversarial_2016, donahue2019large}. 
Intuitively, this originates from the optimization-induced organization of the GANs' latent space where nearby points lead to images that are semantically similar \citep{brock_large_2019}.
This smoothness generalizes when interpolating between distant points in the latent space:~generated samples exhibit smooth transitions from one sample to another, creating new objects that can combine features from multiple distinct objects \citep{berthelot_understanding_2018, brock_large_2019}. 
Exploiting this learned structure, several models invert the generative process of GANs and demonstrate that their latent space contains semantic representations that can be useful to perform downstream tasks \citep{makhzani2015adversarial, dumoulin_adversarially_2017, donahue2019large}.

\medskip
\begin{tcolorbox}
	\textbf{Box 2 - Generative adversarial networks}
	
	Generative Adversarial Networks  \citep[GANs,][]{goodfellow2014generative} introduce a generator $G$ that generates data samples from noise, and a binary classifier, or discriminator ($D$), that distinguishes these generated samples from real data. The generator $G$ is trained to fool the discriminator $D$ into believing that generated samples are real by creating samples that belong to the data distribution. For a sample from the data distribution $\x \sim p(\x)$ and a noise vector sampled from the prior distribution, e.g., $p(\z) \sim \mathcal{N}(0,I)$, the objective of the discriminator $D$ is to {\it minimize} the loss
	   \begin{align}
	 		\mathcal{L}_{\text{adv}} = -  \log D(\x) - \log ( 1-D(G(\z))) \, ,
	 		\label{eqn:GANs}
	   \end{align}
	while the objective of the generator $G$ is to {\it maximize} this loss. This equation defines the cross-entropy loss for a binary classifier ($D$) with a sigmoid output, where the label is $1$ for all data samples $\x$, and $0$ for all generated samples $G(\z)$. Thus, the discriminator improves its ability to discern real from generated samples, while the generator improves the quality of its generated samples so it can fool the (improved) discriminator. After sufficiently many training steps, the generator is able to generate realistic samples, even for complex datasets containing high-resolution images \citep{radford_unsupervised_2015, karras_style-based_2018, brock_large_2019}.
\end{tcolorbox}

\subsection{Adversarial dreaming} 

\begin{figure}[t!]
   \centering
   \includegraphics[width=1.0\textwidth]{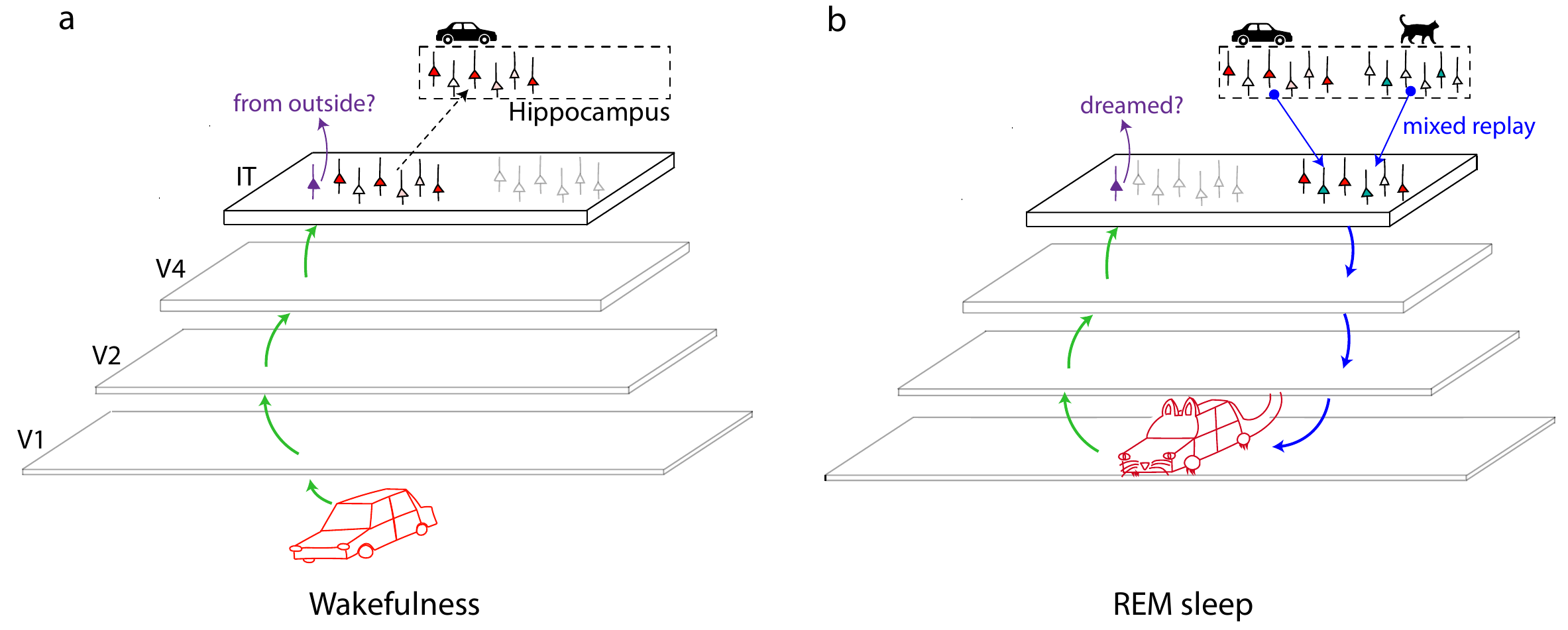}
   \caption{{\bf Learning representations via adversarial dreaming.} 
   \textbf{(a)} During wakefulness, external stimuli are processed from V1 to IT cortex along feedforward pathways (green).
   These learn to recognize the induced early sensory activity as coming from outside (purple neuron).
   Simultaneously, latent representations are stored in the hippocampus.
   \textbf{(b)} During REM sleep several independent memories are replayed from the hippocampus and combined in high-level areas. Feedback pathways (blue) map this latent activity to early sensory areas where virtual experiences (dreams) are generated.
   Following the principles of adversarial learning, feedforward pathways (green) learn to distinguish virtual from stimulus-evoked low-level activities, while feedback pathways improve the generative process to make this distinction harder.
   \label{fig:adversarial}
   }
\end{figure}

Adversarial learning principles have been hypothesized to allow the brain to learn semantic representations from virtual experiences, such as creative dreams, typically occurring during REM sleep \citep{deperrois_learning_2022}. 
In this study, the authors\footnote{when referring to \citet{deperrois_learning_2022}, the authors, although identical to the present paper, are referred to in third person } propose a cortical architecture with a feedback pathway that generates activity in early sensory cortex from high-level representations. 
Additionally, they introduce a feedforward pathway that determines whether activity in lower sensory cortices is externally driven or internally generated.
Feedforward pathways thus assume the role of the discriminator in GANs and are additionally shaped through predictive learning
, being simultaneously trained to infer latent representations from low-level activities (\cref{fig:adversarial}a).
Latent representations inferred during wakefulness are stored in a simple hippocampus model allowing storage and replay. 
When a hippocampal memory is retrieved, feedback pathways are reactivated and generate the associated sensory input.

Learning in this model is organized across three different physiological phases, wakefulness, non-rapid eye movement (NREM) and rapid eye movement (REM) sleep, each characterized by a different objective.
During REM sleep, two different representations from previously observed stimuli are retrieved and together with cortical background activity \citep{spano_dreaming_2020} generate a creative dream through feedback pathways. 
These dreams thus contain elements from both stored memories (\cref{fig:adversarial}b). 
To improve the realism of these virtual experiences, feedback pathways are trained to adversarially fool the FF discriminator into believing that the activity in early sensory areas is externally driven. 
This process defines ``adversarial dreaming''.
Formally, adversarial dreaming is minimizing the classical objective functions of GANs (Box 2) during Wake and REM phases via synaptic weight changes implementing stochastic gradient descent.

The results from this model suggest that REM creative dreams, generated through adversarial dreaming, become more realistic over learning, but still remain different and novel as compared to external sensory inputs \citep{deperrois_learning_2022}, in line with dream phenomenology \citep{nir_dreaming_2010, scarpelli2019mental}. 
Crucially, generating these virtual experiences through both memory combinations and adversarial learning improves the quality of the learned cortical representations.
Indeed, the authors show that object categories can easily be extracted from the latent activity using a linear classifier.
Additionally, they demonstrate that this ability is significantly impaired when they artificially inhibit REM sleep during training.
The authors thus conclude that creative dreams are a key ingredient for the acquisition of semantic latent representations.

\subsection{Neuronal and behavioral correlates }

The principle of adversarial dreaming leads to neuronal and behavioral consequences that can be investigated experimentally. 

A central feature of the framework is that creative dreams during offline states, such as REM sleep, are crucial for the emergence of organized cortical representations. 
This could be tested by recording neuronal population activity in high-level areas using multielectrode arrays. 
From these recordings, one could quantify how well neuronal representations separate object categories, either by training a linear classifier on these representations \citep{hung_fast_2005} or by computing the representation dissimilarity matrices between stimuli \citep{yamins2014performance}. 
We expect that in subjects who are chronically deprived of REM sleep, such as with antidepressant drugs \citep{palagini2013rem}, optogenetic inhibition \citep{boyce_causal_2016, aime2022paradoxical}, or that lack the ability to form mental images \citep[aphantasia,][]{zeman2015lives, pearson_human_2019}, representations are less semantically organized than for control subjects. 
Behaviorally, this would translate as a slower learning speed of novel object classification tasks.

Considering the similarities between mental imagery, imagination and dreaming \citep{kahan1997similarities, llewellyn_crossing_2016, pearson_human_2019}, one could use mental imagery as a practical alternative to dreaming for studying the impact on learning and representation of novel objects in humans.
A potential experiment would involve asking human subjects to classify novel 3D objects and monitoring their learning progress.
Human subjects could be asked to perform mental imagery training sessions following the presentation of novel objects, for instance by mentally rotating them.
We predict that participants who performed these mental tasks would perform better at categorizing these novel objects than the control participants.

Furthermore, in adversarial dreaming, internal activity in early sensory areas becomes more similar to evoked activity over the course of learning, which suggests that dreams should become more realistic with age. 
This correlates with dream reports over different stages of life, that are initially unstructured and plain, and gradually become more meaningful, narrative and less bizarre \citep{nir_dreaming_2010, scarpelli2019mental}. 
According to the theory, this may reflect that older persons know more about the structure of the world and its limitations, and thus become more conservative and less prone to exploration, reducing their capacity to learn new concepts. 
On a neuronal level, this corresponds to an increasing similarity between stimulus evoked and REM generated activity in lower sensory areas. 
In this line, previous work has demonstrated that spontaneous activity, potentially driven by creative daydreaming, indeed becomes more similar to evoked activity in ferret visual cortex over the course of development \citep{berkes2011spontaneous}. 

Finally, in terms of cortical structure, adversarial dreaming predicts a functional organization into two effectively separate feedforward and feedback streams. 
If the information is not forced to go up and down the whole hierarchy, shortcuts between higher cortical areas will prevent lower cortical areas to learn useful features. 
Even though cross-projections between feedforward and feedback pathways are observed experimentally \citep{gilbert_top-down_2013}, adversarial dreaming predicts that those are effectively gated off during essential periods for organizing neuronal representations. 

\subsection{Creativity and adversarial dreaming }

As a consequence of adversarial dreaming, new virtual experiences can be generated by randomly combining different memories (Fig.~\ref{fig:adversarial}b).
This thus leads to the generation of low-level activities that are unlikely to have been evoked by previously experienced stimuli, but that nevertheless may be part of the external world.
While the main focus of this article is to suggest a role of virtual experiences on learning, such a phenomenon suggests two additional functional benefits that are important to mention. 

First, by learning to encode these novel experiences, the system prepares for a future where these imagined sensations are encountered in the wild, such as simulating a dangerous situation offline to escape from it faster when it actually occurs \citep[cf.][]{hobson_rem_2009,llewellyn_dream_2016}. 
Furthermore, generating ``semantic superpositions'' and exposing the feedforward pathways to these may equip the agent with the ability to quickly recognize new stimuli as a composition of known components, making its reaction to them significantly simpler, such as an electric bike leveraging our knowledge about engines and bikes.  
In a behavioral experiment, one could investigate whether participants viewing novel stimuli composed of known parts identify their related categories faster after REM sleep. 

Second, novel adversarially generated experiences could provide an unexpected solution to a specific problem the agent is facing.
During REM sleep, the agent may hence experience an ``insight'' suggesting how to solve a complex problem \citep[also see][]{friston2017active}, such as the Benzene structure that was discovered through a dream by Kekul\'e \citep{mazzarello_what_2000}. 
In this line, generative models are now used in the field of drug discovery to circumvent the limitation of traditional approaches relying upon domain knowledge from physics and chemistry to construct synthesis rules. 
In particular, GAN-based frameworks such as adversarial autoencoders have been used to extend the search of possible molecules for drug design, generating compounds with desired molecular properties \citep{guimaraes2017objective, de2018molgan, blanchard2021using}. 
Naturally, not all creative combinations experienced during dreaming are useful, and their usefulness is ultimately determined by how compatible they are with the actual external world.
This suggests that additional steps may be necessary, such as testing experimentally the  existence of suggested compounds.

More broadly, creative dreaming closely relates to concepts of how to trigger creative thoughts \citep{llewellyn_crossing_2016}:~After being intensively exposed to a certain topic, one needs periods of rest, or ``incubation'' periods, to freely let the generator produce samples. 
Experimentally, this could be tested by evaluating the performance of participants at a creative synthesis task \citep{finke1996creative, palmiero_domain-specificity_2015}, consisting of combining different visual patterns into a new, potentially useful object. 
Subjects that have chronically impaired REM sleep would be less likely to synthesize useful/realistic objects \citep{giancola2022relationships}.
The adversarial dreaming framework thus expands the predictive processing view to the offline generation of creative, virtual experiences that could facilitate the acquisition of semantic representations. We next introduce another unsupervised learning principle that offline states could leverage.

%%%%%%%%%%%%%%%%%%%%%%%%%%%%%%%%%%%%

% CONTRASTIVE LEARNING 

%%%%%%%%%%%%%%%%%%%%%%%%%%%%%%%%%%%%

\section{Learning representations by contrasting sensory experiences}
   \subsection{Principles of contrastive learning}
   
Ultimately, the idea of semantic latent representations is to have similar latent neuronal responses to semantically similar stimuli \citep{dicarlo_how_2012}.
Instead of learning representations implicitly via generative models, one could directly train feedforward pathways to map semantically similar inputs to similar latent representations, and dissimilar inputs to different regions of the latent space \citep[e.g.,][]{le-khac_contrastive_2020}.
In this context, one often refers to similar ("positive") examples as being ``pulled together'' and dissimilar ("negative") examples as being ``pushed apart'' from each other during the training process. 
   
How are positive examples obtained during training, before the network has had the chance to organize its representations? 
Typically, positive examples are created by transforming an existing sample through so-called data augmentations, consisting of cropping, color distorting, or blurring the sample \citep{chen_simple_2020}.
Through this transformation process, the input remains semantically similar to the original input, while its sensory structure can be vastly different.

Negative examples serve to prevent trivial solutions, such as mapping all samples to the same latent vector, often referred to as ``representational collapse'' \citep{le-khac_contrastive_2020, bardes2021vicreg} .
For a given sample, all the other samples from the data set are typically considered negative examples.
Even though this broad definition includes samples from the same category, which can not be excluded in the absence of labels, the majority of negative samples will come from a different category for typical datasets.
Note that recent work suggests that negative examples may not be required for efficient contrastive learning.
Alternative methods include ensuring that representations are variable enough \citep{bardes2021vicreg} or breaking the symmetry between projections of positive examples \citep{grill_bootstrap_2020, chen_exploring_2021}.
Through simple yet effective principles, contrastive learning led to models that are currently state-of-the-art at learning semantic representations useful for downstream tasks in an unsupervised manner \citep{liu_self-supervised_2021, ericsson_selfsupervised_2022}.
   
   \medskip
    \begin{tcolorbox}
	\textbf{Box 3 - Contrastive learning }
	
	Contrastive learning algorithms use an encoder that is trained to compare (latent) representations of data samples. These representations are shaped by pulling together representations of semantically similar inputs and pushing apart those of dissimilar inputs \citep{jaiswal_survey_2020, le-khac_contrastive_2020}. Similar (positive) examples are usually obtained by applying a series of (semantically preserving) data augmentations such as cropping, resizing, blur, color distortion to a given sample \citep{chen_simple_2020}, and negative examples are simply other samples from the dataset. This comparison can be learned with a loss function $\mathcal{L}_{\text{contr}}$ defined on a single positive pair  $(i,j)$ and a large number of negative pairs $(i,k)_{k \neq i}$  such as: 
    \begin{align}
	   \mathcal{L}_{\text{contr}} = - \log \frac{\exp (\text{sim}(\z_i, \z_j)/\tau)}{\sum_{k=1, k \neq i}^{2N}  \exp (\text{sim}(\z_i, \z_k)/\tau ) },\label{eq:contrastive_objective}
    \end{align}
    where where $N$ is the number of examples in a minibatch ($2N$ because all examples are augmented), $\z_i$ is the representation of the sample $k$,  $\text{sim} (\boldsymbol{u}, \boldsymbol{v}) = \boldsymbol{u}^T \boldsymbol{v} / ( \|\boldsymbol{u} \| \, \|\boldsymbol{v} \| )$ denotes the dot product between $l_2$ normalized $\boldsymbol{u}$ and $\boldsymbol{v}$ and $\tau$ denotes the temperature parameter \citep{chen_simple_2020}, i.e., this loss function computes the cosine of the angle between $\boldsymbol{u}$ and $\boldsymbol{v}$. Through this learning objective, the network aims to reduce the distance between the representations of positive pairs ($\z_i, \z_j$) and increase the distance between the representations of negative pairs $(\z_i, \z_k)_{k \neq i}$. 
	\end{tcolorbox}

\subsection{Contrastive dreaming}
   
\begin{figure}[t!]
   \centering
   \includegraphics[width=1.0\textwidth]{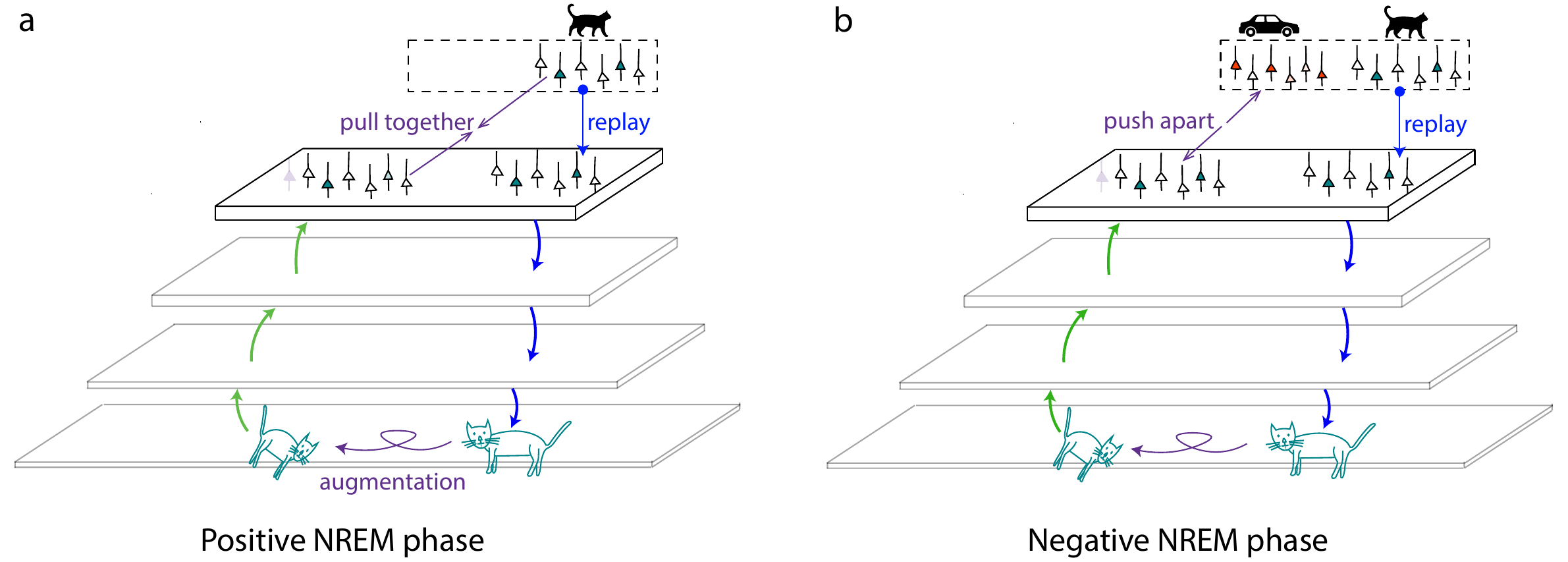}
   \caption{{\bf Learning representations via contrastive dreaming.} 
   \textbf{(a)} During NREM single memories are replayed and generate activities in V1, which are modified in a semantically preserving way, e.g., rotating or squeezing the input. Following the principles of contrastive learning, feedforward pathways learn to map this ``augmented input'' to the same latent representations as the initially replayed hippocampal representation.
   \textbf{(b)} To implement the contrastive step during NREM, feedforward pathways learn to push apart the inferred representation to a different hippocampal memory, serving as a negative example.
   \label{fig:contrastive}
   } 
\end{figure}

Contrastive learning principles may be leveraged by the brain to enhance and robustify neuronal representations during imagination and dreaming.
Just like for adversarial dreaming, the generative model learned by feedback pathways from the prediction of sensory inputs during wakefulness can be leveraged for learning during sleep.
In contrastive dreaming, only a single hippocampal memory serves as the basis for subsequent generation of activity in early sensory cortex. 
Instead of combining multiple stored memories, the virtual experiences thus represent previously observed sensory inputs that are altered by a series of augmentations. 
These augmentations need to be strong enough to change the low-level details of the virtual experience, but not so strong as to change its semantic content, e.g., adding noise to, blurring, cropping, rotating or distorting an image (\cref{fig:contrastive}a). 
These augmentations could be applied by leveraging an additional cortical module, or by directly influencing generation through modulation of feedback pathways at different hierarchical levels \citep{karras_style-based_2018,wybo2023nmda}.
The goal of feedforward pathways then consists of mapping this altered input to its original hippocampal representation, thus pulling together positive pairs (\cref{fig:contrastive}a).

Negative examples are provided by older memories, with feedforward pathways learning to map the augmented experience away from these (\cref{fig:contrastive}b).
Cortically, this could occur after the positive phase, by maintaining the inferred latent representation and comparing it to other hippocampal memories.

This approach was partly explored by \citet{deperrois_learning_2022}. During the NREM phase of the model, virtual experiences generated from single hippocampal memories were partly occluded.
This process made the feedforward network more robust to similar perturbations during perception. 
We hypothesize that by extending the model to additional augmentations, and contrasting it with negative examples (\cref{fig:contrastive}b), such a phase could further improve the semantic organization of the model's latent space. 
In summary, we propose that the efficiency of contrastive learning objectives can be exploited by offline states through contrastive dreams of previous experiences.

\subsection{Neuronal and behavioral correlates}

The contrastive dreaming framework can be experimentally investigated. 
First, it makes predictions about dream phenomenology.
One can assess the diversity of internally generated experiences by waking up sleeping participants at different physiological stages, such as NREM, REM, hypnagogic or day-dreaming states \citep{waters_what_2016}, and asking them to report the content from their dreams, or possibly by directly communicating with them while dreaming \citep{konkoly2021real}. 
We predict that dreams reported from the hypothesized ``contrastive'' states, such as within sharp-wave ripples during NREM sleep \citep{kudrimoti_reactivation_1999}, tend to contain individual previous experiences. 
Depending on the detail of the dream reports, they may even reveal the suggested augmentations, for example in the form of distorted colors or reversed directions. 
In contrast, dreams from adversarial dreaming during REM sleep would be dissimilar to previous experiences but rather combine diverse elements from them \citep{fogel2018novel}.
These predictions are line with experimental data showing that NREM dream reports have more episodic memory sources \citep{baylor_2001_memory} and exhibit less complexity \citep{martin2020structural}. 

Second, contrastive dreaming makes prediction at the neuronal level. 
During wakefulness, as sensory inputs that are nearby in time usually involve the same object under different views \citep{illing_local_2021}, one could expect that over the course of development, activities in high sensory areas of the ventral stream of the visual cortex become increasingly stable during the observation of a moving object.
This could be measured by the representation dissimilarity matrix approach \citep{yamins2014performance} tracking high-level activities over time. 
We would thus expect impaired NREM sleep to lead to more variable neuronal representations. 
More generally, the contrastive dreaming principle predicts that while activities from lower areas are very different across stimulus categories and augmentations, high-level activities should become robust against augmentations but remain sensitive to stimulus categories. 
Additionally, one could compare low-level and high-level activities during NREM sleep.
We predict that while high-level activities resemble waking activities closely due to hippocampal replay, low-level activities vary significantly due to augmentations (\cref{fig:contrastive}a).

%%%%%%%%%%%%%%%%%%%%%%%%%%%%%%%%%%%%

% CONCLUSION  

%%%%%%%%%%%%%%%%%%%%%%%%%%%%%%%%%%%%

\section{Learning beyond the shackles of direct experiences}
\subsection{Summary}

To explain the emergence of semantic neuronal representations in an autonomous, unsupervised manner, influential neuroscientific theories suggest that the brain minimizes prediction errors between its expectations and stimulus-evoked activities \citep{rao_predictive_1999,friston_theory_2005,millidge_predictive_2022,mikulasch2022error}.
Models emerging from these frameworks are successful at describing various properties of cortical networks and can solve complex computational tasks.
However, the rich, sometimes bizarre world of non-sensory related phenomena our brains experience on a nightly basis appear only to reduce the complexity of the generative model.
To complement predictive processing theories, here we discussed two computational frameworks through which brains can benefit even more from their internally generated virtual experiences.

First, adversarial dreaming combines several stored memories with cortical noise and pits feedback and feedforward pathways against each other in a creative game of generating and discriminating low-level activities. This process thereby implicitly learns an organized latent structure. 
Second, contrastive dreaming explicitly trains feedforward networks to map semantically similar inputs to similar high-level cortical representations by dreaming up previously observed inputs with semantically preserving augmentations.
Both principles are compatible with the bidirectional architecture of sensory cortices and could be implemented in network models relying on biologically plausible credit assignment algorithms and learning rules \citep{richards_2019, lillicrap_backpropagation_2020}.

While our principles primarily pertain to dreaming, we anticipate their applicability to other forms of spontaneous, virtual experiences like mental imagery \citep{pearson_human_2019}, meditation \citep{cooper2022beyond}, and spontaneous thoughts \citep{mildner2019spontaneous}. The key distinction lies in the nature of these experiences: adversarial dreaming involves the creative recombination of memory elements, whereas contrastive dreaming reenacts "augmented" past experiences. We propose that virtual experiences during dreamlike states generally align with one of these two mechanisms.

\subsection{Outlook: learning from predictive, adversarial and contrastive principles}

Despite their algorithmic differences, the three presented learning principles can be implemented by the same cortical architecture. They however require different physiological phases, in line with previous theories \citep{hinton_wake-sleep_1995, giuditta1998, hobson_waking_2012, lewis_how_2018}.

An interesting direction would be to explore whether the combined optimization of different learning objectives could have a synergistic effect on the acquisition of semantic representations. 
While the combination of predictive and adversarial learning has been previously explored \citep{makhzani2015adversarial, brock_neural_2017, ulyanov_it_2017}, the benefits of combining contrastive and adversarial principles remain to be elucidated \citep[but see][]{chen_self-supervised_2019, deperrois_learning_2022}.

Finally, humans develop under {\it some} supervision, for example in the form of explicit verbal instructions about object categories.
It is hence natural to explore the combination of the unsupervised learning principles described so far with sparse labels to further improve the learned latent structure \citep[see also][]{deperrois_learning_2022}.

Since these principles are in many ways complementary, experimentally the influence of each may be challenging to tease apart. 
However, despite their similarities, they have different functional goals. 
Predictive processing allows the brain to predict upcoming stimuli, adversarial dreaming aims to prepare the brain for previously unobserved stimuli, and contrastive dreaming aims to make latent representation invariant to irrelevant factors of variation.
Through carefully designed experiments, for instance analyzing the individual effects of NREM and REM sleep on cortical dynamics \citep{tamaki2020complementary}, or by analyzing the effect of different learning tasks on NREM and REM activity patterns \citep{fogel2006learning, fogel2007dissociable} their different functional goals may hence be exploited to tease apart their influences on neuronal representations.

\subsection{Relation to previous work}

\citet{gershman_generative_2019} proposed an adversarial framework for brain function in view of psychological and neural evidence.
In particular, he discusses the consequences of a dysfunctioning discriminator on the perception of hallucinations, leading to potential delusions observed in mental disorders. 
The framework discussed here may serve as a suggestion for implementing a mechanistic model of these ideas and further elucidate the consequences of dysfunction in specific modules.

Previous work suggested an alternative explanation for the creative aspect of dreams during REM sleep.
The pioneer activation-synthesis theory from \citet{hobson_brain_1977} suggests that REM dreams result from the brain ``making the best of a bad job in producing even partially coherent dream imagery from the relatively noisy signals sent up to it from the brain stem''.
Adversarial dreaming provides a concrete instantiation of this idea by forming coherent REM dreams from incoherent signals. From replaying a random mixture of episodes out of the hippocampus to the cortex, the discriminator network provides the feedback to increase the realism of the generated dream imagery. 
Other authors attribute this creative phenomenon to a shift of topographical neural activity towards the Default Mode Network \citep{domhoff2015dreaming}, encouraging external inputs from wakefulness to be integrated with internally generated imagery, jointly manifesting as bizarre dream content \citep{northoff2023topographic}.

Generative modeling's early developments, notably the Wake-Sleep algorithm \citep{hinton_wake-sleep_1995}, previously emphasized the importance of offline processes in optimizing latent representations.
First, the sleep phase in the Wake-Sleep algorithm, while conceptually different, is algorithmically akin to the positive phase of 'contrastive dreaming,' where generated inputs are aligned with their originating latent activities via the feedforward network.
Second, this algorithm introduced a bidirectional structure of cortical projections, where feedforward pathways encode sensory inputs and feedback pathways generate sensory predictions or fictive inputs.
This concept later influenced the development of variational autoencoders \citep{kingma2013autoencoding}; their biological plausibility was recently examined in \citep{marino_predictive_2022}.
The introduced framework here also leverages a bidirectional organization necessary to implement both adversarial and contrastive dreaming paradigms.
In this view, backward projections serve a dual role of making predictions during wakefulness, while generating virtual experiences during offline states.

Previous work on predictive processing suggests that its principles extend beyond waking experiences \citep{hobson_waking_2012, hobson_virtual_2014}.
This theory rests on the free-energy principle, which formulates processing and learning as a variational optimization problem.
Intuitively speaking, a good model should provide a good explanation of observed data ("model accuracy"), while maintaining a minimal set of assumptions ("model complexity"), together maximizing "model evidence" \citep{friston_free-energy_2010}.
Accordingly, wakefulness provides an opportunity to optimize both of these components, while dreams, or more generally offline states, specifically allow for the reduction of model complexity.
Indeed, such ideas have been successfully employed for machine learning models \citep{ponnapalli1999formal, simoncelli2001natural, williams1995bayesian} and suggested to provide a functional explanation for synaptic homeostasis during sleep \citep{tononi_sleep_2014}: minimizing the brain's model complexity may improve generalization abilities.

Similar to predictive processing, adversarial dreaming also aims to maximize model evidence, though implicitly with the help of feedforward pathways, rather than explicitly \citep{huszar2017variational}.
Nevertheless, this similarity in spirit suggests that adversarial dreaming too could benefit from the reduction of complexity as suggested by \citet{hobson_virtual_2014}.
Vice versa, predictive processing could benefit from the ability of feedforward pathways being able to distinguish between internally generated and externally driven activities in sensory cortex, learned via adversarial dreaming.
While predictive processing learns semantic representations implicitly, contrastive dreaming explicitly optimizes these behaviorally relevant variables.
Nevertheless, the neuronal representations emerging from contrastive learning may also help generative models to maximize model evidence.
These observations suggest an intimate relation between these theories, jointly highlighting the importance of virtual, non-sensory, experiences.

Two recent studies suggested how the brain could benefit from constrastive objectives.
In \citet{illing_local_2021}, the authors propose that positive examples are obtained from the observation of a moving object, while negative examples appear through saccades towards new objects.
% They notably show that these objectives can be optimized at a local-level via Hebbian learning rules, making them more biologically plausible. 
%
In contrast, \citet{halvagal_2022} argue that the brain does not need negative examples, as long as latent activities are encouraged to remain sufficiently variable \citep[through variance maximization,][]{bardes2021vicreg}. 
Through this mechanism, networks learn invariant representations for stimulus features which change slowly in time. 
A downside from these models is that positive examples, and thus augmented inputs, are assumed to be obtained through the observation of moving objects, leading to limited augmentation. 
However, a series of strong augmentations are needed to obtain strong semantic representations via contrastive learning objectives \citep{chen_simple_2020}.
To avoid interference of such strong augmentations with perception, 
hosting them during offline states as suggested by the contrastive dreaming principle thus provides a beneficial alternative.

\subsection{Conclusion: the necessity of virtual experiences for learning}

We proposed that essential processes shaping our cortical function arise from brains generating virtual experiences during sleep. 
Learning from an imagined world may thus be just as important as learning from sensations. 
This view significantly expands our perspective on perception and learning.
Do we need to be constantly focused on our sensorium to learn optimally, or can we finally justify sometimes having our heads in the clouds?

\paragraph{Acknowledgements}
This work has received funding from the European Union $7$th Framework Programme under grant agreement 604102 (HBP), the Horizon 2020 Framework Programme under grant agreements 720270, 785907 and 945539 (HBP), the Swiss National Science Foundation (Sinergia grant CRSII5-180316), the Interfaculty Research Cooperation (IRC) `Decoding Sleep' of the University of Bern, and the Manfred St{\"a}rk Foundation.

\bibliographystyle{apalike}
\bibliography{references_review.bib}

\end{document}